\documentclass[
 reprint,
 superscriptaddress,
 showpacs
 preprintnumbers,
 amsmath
 amssymb,
 aps,
 prx,
 floatfix,
]{revtex4-2}

\DeclareMathAlphabet{\mathsfit}{T1}{\sfdefault}{\mddefault}{\sldefault}
\SetMathAlphabet{\mathsfit}{bold}{T1}{\sfdefault}{\bfdefault}{\sldefault}

\usepackage{graphicx}
\graphicspath{{fig/}}
\usepackage{dcolumn}
\usepackage{bm}
\usepackage{makecell}
\usepackage{braket}
\usepackage{siunitx}
\usepackage{color}
\usepackage{array}
\usepackage{amsmath}
\usepackage{xr}
\usepackage{upgreek}
\usepackage[colorlinks]{hyperref}
\usepackage[capitalize]{cleveref}
\usepackage[caption=false]{subfig}
\usepackage{multirow}

\begin{document} 

\title{Learning-based Calibration of Flux Crosstalk in Transmon Qubit Arrays}

\def\RLEaffil{Research Laboratory of Electronics, Massachusetts Institute of Technology, Cambridge, MA 02139, USA}
\def\LLaffil{MIT Lincoln Laboratory, Lexington, MA 02421, USA}
\def\Physaffil{Department of Physics, Massachusetts Institute of Technology, Cambridge, MA 02139, USA}
\def\EECSaffil{Department of Electrical Engineering and Computer Science, Massachusetts Institute of Technology, Cambridge, MA 02139, USA}
\def\Wellesleyaffil{Department of Physics, Wellesley College, Wellesley, MA 02481, USA}
\def\Infineonaffil{Infineon Technologies, Munich, Germany}

\author{Cora~N.~Barrett}
\thanks{These authors contributed equally to this work.}
\affiliation{\RLEaffil} 
\affiliation{\Wellesleyaffil}

\author{Amir~H.~Karamlou}
\thanks{These authors contributed equally to this work.}
\email{karamlou@mit.edu}
\affiliation{\RLEaffil}
\affiliation{\EECSaffil}

\author{Sarah~E.~Muschinske}
\affiliation{\RLEaffil}
\affiliation{\EECSaffil}

\author{Ilan~T.~Rosen}
\affiliation{\RLEaffil}

\author{Jochen~Braumüller}
\affiliation{\RLEaffil}

\author{Rabindra~Das}
\affiliation{\LLaffil}

\author{David~K.~Kim}
\affiliation{\LLaffil}

\author{Bethany~M.~Niedzielski}
\affiliation{\LLaffil}

\author{Meghan~Schuldt}
\affiliation{\LLaffil}

\author{Kyle~Serniak}
\affiliation{\RLEaffil}
\affiliation{\LLaffil}

\author{Mollie~E.~Schwartz}
\affiliation{\LLaffil}

\author{Jonilyn~L.~Yoder}
\affiliation{\LLaffil}

\author{Terry~P.~Orlando}
\affiliation{\RLEaffil}
\affiliation{\EECSaffil}

\author{Simon~Gustavsson}
\affiliation{\RLEaffil}

\author{Jeffrey~A.~Grover}
\affiliation{\RLEaffil}

\author{William~D.~Oliver}
\email{william.oliver@mit.edu}
\affiliation{\RLEaffil}
\affiliation{\EECSaffil}
\affiliation{\LLaffil}
\affiliation{\Physaffil}

\date{\today}

\begin{abstract}

Superconducting quantum processors comprising flux-tunable data and coupler qubits are a promising platform for quantum computation.
However, magnetic flux crosstalk between the flux-control lines and the constituent qubits impedes precision control of qubit frequencies, presenting a challenge to scaling this platform.
In order to implement high-fidelity digital and analog quantum operations, one must characterize the flux crosstalk and compensate for it.
In this work, we introduce a learning-based calibration protocol and demonstrate its experimental performance by calibrating an array of 16 flux-tunable transmon qubits.
To demonstrate the extensibility of our protocol, we simulate the crosstalk matrix learning procedure for larger arrays of transmon qubits.
We observe an empirically linear scaling in calibration time with system size while maintaining a median qubit frequency error below $\SI{300}{kHz}$.

\end{abstract}

\maketitle

\section{Introduction}

Superconducting quantum processors comprising flux-tunable transmon qubit arrays are at the forefront of contemporary digital quantum computation~\cite{arute_2019, acharya_2022} and analog quantum simulation and emulation~\cite{zhang_2023, karamlou_quantum_2022, braumuller_probing_2022}.
Flux-tunable qubits enable controllable, strong qubit-qubit interactions and high-fidelity two-qubit gates~\cite{yan_2018_coupler, sung_2021} in many-qubit systems by reducing parasitic couplings and qubit frequency crowding.
A central requirement for operating such quantum devices is the accurate and precise frequency control of each tunable element using local flux lines, which can be used to tune each qubit individually. 
While a current applied to a particular flux-line antenna is designed to address only one qubit or coupler, there are at least two mechanisms by which flux may couple to additional elements.
One is through the direct, unwanted inductive coupling from the antenna to other qubits and couplers.
A second is via the send and return path of the applied current, which may similarly induce unwanted magnetic flux in other qubits as it traverses the flux line and ground plane. 
The net effect is commonly referred to as flux crosstalk.
Crosstalk can be characterized and compensated by measuring the response of each qubit to the flux generated by every flux line independently and presuming linear superposition, or by using an iterative approach~\cite{dai_calibration_2021} and optimization~\cite{dai_periodicity_2022} in systems with a non-linear response.

Machine learning techniques have been applied extensively to calibrating quantum dots~\cite{Darulova_tuning_2020,van_Esbroeck_unsupervised_2020, Lennon_measuring_2019, Moon_automatic_2020, Severin_tuning_2021, Ziegler_autotuning_2022, Ziegler_extraction_2023, Zwolak_automation_2021, Zwolak_doubledot_2020}.
For superconducting quantum processors, machine learning approaches have been used, for example, for automated recalibration of system parameters~\cite{Kelly_optimus_2018}, optimization of qubit frequency layouts~\cite{klimov_snake_2020}, discrimination of qubit states~\cite{magesan_ml_2015, martinez_hmm_2020, navarathna_nn_2021, duan_nn_2021, lienhard_nn_2022}, and calibration of single- and two-qubit gates~\cite{Baum_reinforcement_2021, Wittler_toolset_2021}.
In this work, we demonstrate and analyze a learning-based approach for characterizing flux crosstalk on flux-tunable transmon devices.
In comparison to previous works, our approach does not involve direct measurement of crosstalk matrix elements and requires relatively few measurements.

Flux-tunable transmons comprise two Josephson junctions forming a superconducting quantum interference device (SQUID) in parallel with a shunting capacitor~\cite{koch_2007}.
In this circuit, the Josephson energy $E_J$ is tuned by threading an external magnetic flux $\Phi_\mathrm{ext}$ through the SQUID loop.
The transition frequency between the ground and the first-excited state of the transmon in response to the applied magnetic flux is approximately given by~\cite{koch_2007}
\begin{align}
    & f(\Phi_\mathrm{ext}) \approx \nonumber \\
    & \left(f^{\text{max}} + \frac{E_{C}}{h}\right)\sqrt[4]{d^2 + (1-d^2)\:\text{cos}^2\left(\pi \frac{\Phi_\mathrm{ext}}{\Phi_0}\right) } - \frac{E_{C}}{h}
\label{eq:transmon_spectrum}
\end{align}
where $f^{\text{max}}=(\sqrt{8 E_J E_C}-E_C)/h$ is the maximum qubit frequency, assuming $E_J\gg E_C$, and $E_{C}$ is the transmon charging energy.
The asymmetry parameter $d$ of the SQUID junctions is given by $d = |(E_{J,2}-E_{J,1})/(E_{J,2}+E_{J,1})|$, where $E_{J,1}$ and  $E_{J,2}$ are the Josephson energies of the two SQUID junctions.
The transmon spectrum in Eq.~\ref{eq:transmon_spectrum} provides a formula to estimate the applied magnetic flux required for tuning the qubit to a particular frequency.

In a flux-tunable transmon processor, the magnetic flux is applied by running an electric current through a flux line terminated by an antenna that is near the target qubit and inductively coupled to its SQUID loop.
The current is generated at room temperature either by using an active current source or by using a voltage source outputting voltage $V$ across a series resistance $R$. In either case, it is important to use a ``stiff" current source with high output resistance to ameliorate the impact of temperature-dependent line resistance inside the refrigerator leading to the qubits. 
In this work, we use a voltage source and resistor $\SI{1}{k \ohm}$ to apply a current to each of our flux bias lines.
The magnetic flux can be expressed as $\Phi_\mathrm{ext} = V/V^{\Phi_0}+ \Phi_{\text{offset}}$, where $V^{\Phi_0}$ is the voltage required to tune the qubit by one magnetic flux quantum $\Phi_0$, and $\Phi_{\text{offset}}$ is a flux offset due to magnetic fields produced by vortices trapped in the superconducting ground plane or other non-controllable sources of static magnetic field.

\begin{figure}[ht]
\subfloat{\label{fig:crosstalk_concepts}}
\subfloat{\label{fig:physical_device}}
\subfloat{\label{fig:error_dist}}
\includegraphics{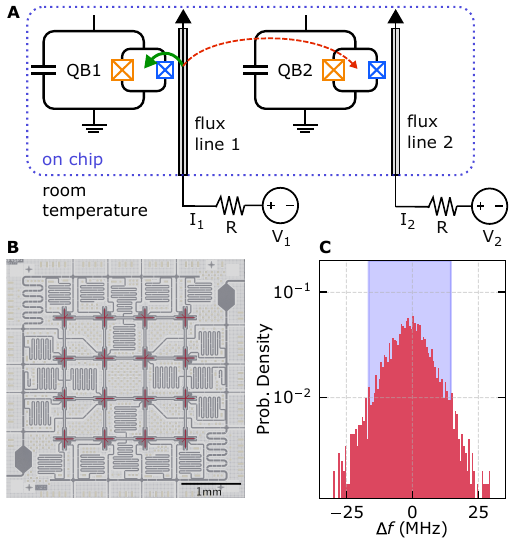}
\caption{
\textbf{The Flux Crosstalk Model}
\textbf{(A)} Flux crosstalk concept for superconducting transmon circuits.
Current $I_1$ passing near the SQUID of qubit $1$ supplies an intended amount of magnetic flux (green arrow).
Due to crosstalk, other qubits on the device may experience an unintended flux from this current (dashed red arrow).
\textbf{(B)} Optical image of the qubit tier of the 16-qubit transmon array fabricated using a 3D-integrated, flip-chip process (see Appendix~\ref{appendix:device} for details of the device). The capacitor pads of the qubits are false-colored maroon.
\textbf{(C)} Experimental distribution of the difference between the measured qubit frequency and the target frequency $\Delta f= f_q - f_{\text{target}}$ without crosstalk correction for 200 target frequency vectors $\vec{f}_{\text{target}}$. For each $\vec{f}_{\text{target}}$, all 16 qubits are simultaneously biased to random frequencies in the region spanned by $\SI{100}{MHz}-\SI{1}{GHz}$ below the maximum qubit frequency, and then each qubit frequency is measured.
The shaded region indicates the $5^{\mathrm{th}}$ to $95^{\mathrm{th}}$ percentiles of the distribution.}
\label{fig:DEV}
\end{figure}

Magnetic flux crosstalk can be treated as a linear process; a vector of voltages $\vec{V}$ applied to the flux lines is related to the magnetic flux $\vec{\Phi}_\mathrm{ext}$ experienced by the qubits by the relation
\begin{equation}
    \vec{\Phi}_\mathrm{ext} = {(\boldsymbol{V}^{\Phi_0})^{-1}\boldsymbol{S}}\vec{V} + \vec{\Phi}_{\text{offset}}\,,
    \label{eq:crosstalk_eq}
\end{equation}
where $\boldsymbol{V}^{\Phi_0}$ is a diagonal matrix with $\boldsymbol{V}^{\Phi_0}_{i,i}$ corresponding to the $V^{\Phi_0}$ of qubit $i$.
${\boldsymbol{S}}$ is the flux crosstalk sensitivity matrix, with $\boldsymbol{S}_{i,j}=\partial V_i /\partial V_j$ representing the voltage response of qubit $i$ to a voltage signal applied to qubit $j$.
In this representation, the diagonal elements of ${\boldsymbol{S}}$ are 1, and characterizing $\boldsymbol{S}$ along with $\vec{\Phi}_\mathrm{offset}$ enables us to compensate for the crosstalk and set the qubit frequencies more precisely. Before this compensation, we observe a spread in frequency error $\Delta f$ as shown in Fig.~\ref{fig:error_dist}.

The flux crosstalk sensitivity matrix for an array of $N$ flux-tunable transmon qubits contains $N^2$ elements.
We note that since the qubit array and routing layouts are not generally symmetric, the matrix $\boldsymbol{S}$ is not guaranteed to be symmetric. In general, $\boldsymbol{S}_{i,j} \neq \boldsymbol{S}_{j,i}$, and so all $N^2$ elements need to be characterized.
Typically, each element $\boldsymbol{S}_{i,j}$ is individually characterized by sweeping voltage $V_j$ targeting qubit $j$ and measuring the response of qubit $i$~\cite{abrams_methods_2019,braumuller_probing_2022,krinner_2022}.
This approach, however, is not extensible for characterizing the flux crosstalk of large transmon qubit arrays (see Appendix~\ref{appendix:direct_measurement}).
The sample is fabricated using a 3D-integrated, flip-chip process~\cite{rosenberg_2017}, where the resonators and the flux lines are located on the interposer tier, and the qubits are located on a separate qubit tier.
This learning-based calibration approach has been employed to calibrate the flux crosstalk on planar as well as flip-chip devices~\cite{karamlou_quantum_2022,braumuller_probing_2022,karamlou_2023}.

\section{Calibration protocol}
\subsection{Learning the flux crosstalk matrix}

We learn the flux crosstalk matrix $\boldsymbol{S}$ by attempting to set our qubits to a target frequency layout, measuring the qubit frequencies as a result of applied external flux, and then using the frequency error to optimize $\boldsymbol{S}$. More specifically, we apply a set of voltages $\vec{V}$, where the $i$-th element of $\vec{V}$ denotes the voltage applied to the flux line targeting qubit $i$, and measure the frequency of qubit $i$ to infer the flux $\Phi_{\mathrm{ext}, i}$ via Eq.~\ref{eq:transmon_spectrum}. Minimizing the difference between the measured flux, $\Phi_{\mathrm{ext}, i}$, and the estimated flux, which depends upon $\vec{V}$ and $\boldsymbol{S}$, optimizes $\boldsymbol{S}$.

\begin{figure}[t]
\subfloat{\label{fig:transmon_spectrum}}
\subfloat{\label{fig:spectroscopy}}
\subfloat{\label{fig:simulation_error}}
\subfloat{\label{fig:matrix_distance}}
\includegraphics{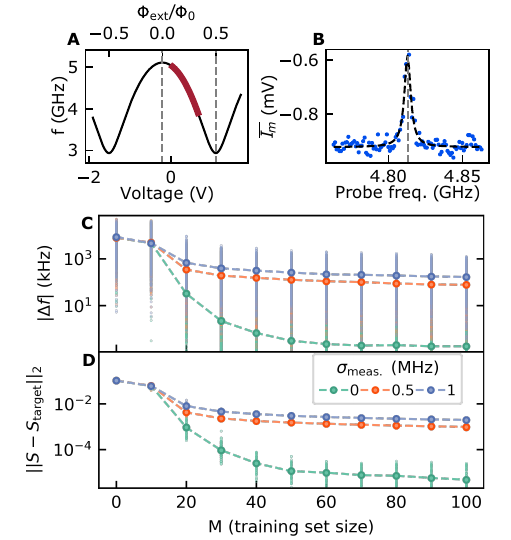}
\caption{
\textbf{Learning protocol simulation} of measurement uncertainty dependence for a 16-qubit array.
\textbf{(A)} Frequency spectrum of the transmon as a function of external flux applied.
We bias the qubits to the maroon region of the spectrum for the flux crosstalk training.
\textbf{(B)} We measure qubit frequencies experimentally via spectroscopy, by measuring the magnitude of the in-phase quadrature of the demodulated signal ($\overline{I_m}$), and extract the qubit frequency from a Lorentzian fit.
\textbf{(C)} Simulated frequency error $|\Delta f|$ and the \textbf{(D)} Euclidean distance between the trained crosstalk matrix $\boldsymbol{S}$ and the target matrix $\boldsymbol{S}_{\text{target}}$ when the matrix is learned in the presence of qubit frequency measurement uncertainty $\sigma_{\mathrm{meas.}} \in \{0,0.5,1\}$~MHz.
The distribution over $100$ different random realizations is shown using small circles, and the median values are shown using large circles. 
Both quantities converge quickly in less than $M=100$ training sets.
The protocol accuracy decreases as the frequency measurement uncertainty increases.}
\label{fig:protocol_simulation}
\end{figure}

We generate $\vec{V}$ by randomly selecting a frequency for each qubit, $\vec{f}$, subject to a few constraints.
First, we require that each qubit frequency falls in the range spanned by approximately $\SI{100}{MHz}-\SI{1}{GHz}$ below its maximum qubit frequency (colloquially referred to as the ``sweet spot'' due to its first-order insensitivity to flux noise), as illustrated in Fig.~\ref{fig:transmon_spectrum}. 
Second, to reduce frequency shifts due to resonant interaction between qubits, we require that neighboring qubits in the array are far detuned ($>\SI{200}{MHz}$), and the detuning between any two qubits is at least $\SI{50}{MHz}$.
The target frequency range includes regions of the transmon spectrum close to the sweet spot, which are less sensitive to changes in flux. This qubit frequency placement is necessary for sufficiently detuning the qubits.
The detuning mitigates frequency shifts due to resonant interaction between qubits (see Appendix~\ref{appendix:device}), enabling us to accurately measure the frequencies of all qubits simultaneously. 

To minimize bias in this quasi-random selection, we randomly permute the order in which target qubit frequencies within $\vec{f}$ are chosen. 
This random permutation ensures that, given the detuning constraints, the qubits with higher sweet-spot frequencies in the latter half of $\vec{f}$ are not regularly placed at the higher frequency end of the training region.
Using the transmon spectrum in Eq.~\ref{eq:transmon_spectrum}, we can calculate the magnetic flux values corresponding to the target frequencies.
An initial guess for $\vec{V}$ can be obtained from the relationship described in Eq.~\ref{eq:crosstalk_eq} by assuming $\boldsymbol{S}=\boldsymbol{I}$ or using an estimate of the matrix.
All elements of $\vec{V}$ will be nonzero since each qubit is biased off of its sweet spot.

We apply $\vec{V}$ and measure the frequency of each qubit. In our experiments, these frequency measurements are performed simultaneously.
We then convert the measured frequencies into the flux experienced by each SQUID, $\vec{\Phi}_{\text{meas}}$.
By repeating this procedure for $M$ iterations, we obtain a set $\{\vec{V}_i, \vec{\Phi}_{\text{meas},i} \}_{i=1:M}$ of input voltages and the resulting measured fluxes experienced by the qubits.
Using this data set, we train the elements of the $k^{\text{th}}$ row of $\boldsymbol{S}$ by minimizing the mean-squared-error cost function
\begin{align}
    & C(\boldsymbol{S}_k) = \nonumber \\ 
    &\frac{1}{M} \sum\limits_{i=1}^M \Bigg|\Bigg| (\vec{\Phi}_{\text{meas},i})_k
    - \Big[ {(\boldsymbol{V}^{\Phi_0}_{k,k})^{-1}\boldsymbol{S}_k} \vec{V}_i+
    (\vec{\Phi}_{\text{offset}})_k \Big]\Bigg|\Bigg|^2\,.
\end{align}

The first term in the cost function sum corresponds to the measured flux on qubit $k$, and the second term corresponds to the estimated flux based on our crosstalk matrix and the applied voltages. 
We minimize $C(\boldsymbol{S}_{k})$ with the L-BFGS gradient descent optimization algorithm~\cite{liu_lbfgs_1989} implemented in \textsf{PyTorch}~\cite{paszke_pytorch_2019} (see Appendix~\ref{appendix:optimizers} for a comparison of different optimizers).
For a visual schematic of the calibration protocol, see Appendix~\ref{appendix:flowchart}.
The minimization of the cost function will converge as the estimated fluxes approach the measured fluxes.

\subsection{Protocol simulation}

We analyze the performance of our protocol for learning the crosstalk matrix on a simulated model of the 16-qubit array of transmon qubits (shown in Fig.~\ref{fig:physical_device}).
Appendix~\ref{appendix:crosstalk_simulation_model} contains further details of the simulation model.
In Fig.~\ref{fig:simulation_error}, we report the error in achieving the desired target frequency $\Delta f = f_q - f_{\text{target}}$ when training the matrix using a training set of size $M$. 

For the simulations, we modeled the flux crosstalk calibration protocol using realistic values for the transmon spectra parameters ($\boldsymbol{V}^{\Phi_0}$, $\Phi_{\text{offset}}$, $f^{\text{max}}$, $E_{C}$, and $d$).
For the optimal crosstalk matrix $\boldsymbol{S}_{\text{target}}$, we used the previously-characterized crosstalk matrix for the 16-qubit array of transmons.
In these simulations, each time a qubit frequency is ``measured,'' we add normally distributed error
with standard deviation $\sigma_{\mathrm{meas.}}$, reflecting frequency-measurement imprecision (Fig.~\ref{fig:spectroscopy}) due to the qubit spectral linewidth arising from dephasing and power broadening, as well as frequency shifts arising from dispersive shifts and residual ZZ couplings~\cite{Schuster_stark_2005}.
We find that $\sigma_{\mathrm{meas.}} \approx \SI{0.5}{MHz}$ most closely reflects our experimental conditions (see Appendix~\ref{appendix:freq_error_budget} for a discussion of frequency measurement error sources). Based on the trend in Fig.~\ref{fig:simulation_error}, we expect that the plateau frequency error $|\Delta f|$ can be reduced by performing finer spectroscopy scans or measuring qubit frequencies via Ramsey interferometry, at the cost of longer overall calibration time.
Another source of frequency measurement error arises due to small dispersive frequency shifts, despite the detuning scheme. We compensate for these detunings by using previously-characterized qubit-qubit couplings, but there could be some error in the calculation of the uncoupled qubit frequency.

We then create a training set of size $M$ and train $\boldsymbol{S}$.
Finally, we validate our learned crosstalk matrix $\boldsymbol{S}$ with a random vector of frequencies $\vec{f}_{\text{target}}$ independent of the training set. 
These random frequency vectors are chosen in the same way as the target frequency vectors used to learn the matrix, i.e. all qubits are biased to $\SI{100}{MHz}-\SI{1}{GHz}$1 off the sweet spot with all qubits sufficiently detuned.
We attempt to set the qubits to these frequencies and record the error $|\vec{f}_q - \vec{f}_{\text{target}}|$, with no error added to the simulated measurement of $\vec{f}_q$.

We study the convergence of the protocol in the presence of frequency measurement errors. 
We observe that in the absence of measurement uncertainty ($\sigma_{\mathrm{meas.}} = 0~\text{MHz}$), the protocol is capable of achieving a median frequency error $|\Delta f|$ on the order of $\SI{1}{kHz}$ with $M=100$ training sets.
In Fig.~\ref{fig:matrix_distance} we consider the Euclidean distance between the trained crosstalk matrix $\boldsymbol{S}$ and the target crosstalk matrix $\boldsymbol{S}_{\text{target}}$: $||\boldsymbol{S}-\boldsymbol{S}_{\text{target}}||_2$.
As the frequency measurement uncertainty increases, the performance of the protocol degrades, leading to a larger $|\Delta f|$ value when reaching the training plateau.
We expect that with the anticipated uncertainty in measuring qubit frequencies in experiments, we should be able to achieve a median frequency error on the order of $\SI{100}{kHz}$.

\section{Experimental Results}
\label{sec:experiment}

Next, we experimentally assess the performance of our protocol by calibrating the static flux-crosstalk matrix for a 16-qubit array of transmons.
Starting from the assumption that $\boldsymbol{S}=\boldsymbol{I}$, we generate a set of $200$ random voltage vectors, apply each voltage vector to the flux lines, and measure the corresponding qubit frequencies simultaneously via spectroscopy.
Despite targeting $>\SI{200}{MHz}$ detuning between neighboring qubits, each qubit experiences a frequency shift due to its interaction with other qubits.
We calculate the uncoupled qubit frequencies from the measured shifted frequencies using the pre-characterized qubit-qubit couplings.
Using Eq.~\ref{eq:transmon_spectrum}, we convert these frequencies into a vector of fluxes experienced by each SQUID loop in the system corresponding to each applied voltage vector.
We use different subsets of the measured set $\{\vec{V}_i, \vec{\Phi}^{\text{meas}}_i \}_{i=1:200}$ to learn the device crosstalk matrix.

\begin{figure}[hbt!]
\subfloat{\label{fig:error_dist_N_10}}
\subfloat{\label{fig:error_dist_N_20}}
\subfloat{\label{fig:error_dist_N_100}}
\subfloat{\label{fig:experimental_scaling}}
\includegraphics[width=1.0\linewidth]{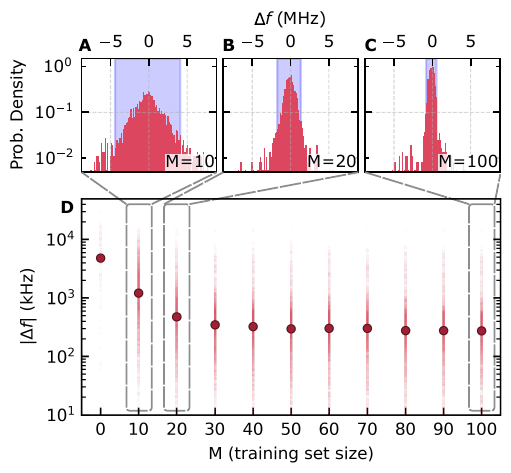}
\caption{
\textbf{Experimental implementation} of the protocol using a 16-qubit array of transmon qubits.
We see the distribution of offset in the measured qubit frequencies from the target values ($\Delta f$) for the crosstalk matrix trained with \textbf{(A)} $M=10$, \textbf{(B)} $M=20$, and \textbf{(C)} $M=100$ number of training sets.
The shaded region indicated the 5$^{\text{th}}$ to 95$^{\text{th}}$ percentile of the distribution.
\textbf{(D)} Error in targeting the qubit frequency $|\Delta f|$ after training the crosstalk matrix with a different number of measurement sets.
We observe that by increasing the training set size $M$, the median error (maroon points) decreases until reaching a plateau of approximately $\SI{290}{kHz}$.}
\label{fig:Exp}
\end{figure}

Prior to training, we selected a validation set of $10$ target frequency vectors. This validation set was generated in the same way as target frequency vectors for training, with the added condition that each qubit is placed at least twice in the upper, middle, and lower regions of the frequency training region. 
For each value of $M$, we randomly selected $20$ different subsets of $\{\vec{V}_i, \vec{\Phi}^{\text{meas}}_i \}$ with size $M$.
With each subset, we learned the crosstalk matrix and then recorded the frequency error in setting the qubits to the validation frequencies.

In Figs.~\ref{fig:error_dist_N_10},~\ref{fig:error_dist_N_20}, and~\ref{fig:error_dist_N_100} we show the qubit frequency deviation from the target validation set value ($\Delta f$) for training sets of size $M=10, \,20, \,\textrm{and} \,100$, respectively.
The shaded region in the figures indicates the 5$^{\text{th}}$ to 95$^{\text{th}}$ percentiles of the distribution.
We observe that by using a larger training set, the distribution of $\Delta f$ becomes narrower.

In our experiments, we notice irregularities in the transmon spectra of four of the qubits, presumably due to two-level-system (TLS) defects coupled to the qubits. The frequency of a qubit coupled to coherent defects shifts, resulting in deviations from the transmon spectrum. In such frequency regions, we experience an error in setting the frequency of just a single qubit.
Therefore, we do not include those qubits in the validation. 
The flux crosstalk for these four qubits can still be learned by excluding the regions of their transmon spectra impacted by TLSs.
We can repeat the same learning-based protocol, biasing each of the four qubits of interest to a defect-free frequency, while applying quasi-random voltages to all other qubits and measuring the frequencies of the four qubits of interest.
The resulting voltages and fluxes can be used to learn the corresponding four rows of $\boldsymbol{S}$.
In Appendix~\ref{appendix:crosstalk_simulation_model}, we report the full learned $\boldsymbol{S}$ for the 16-qubit array.

We demonstrate the experimental scaling of the frequency error $|\Delta f|$ as a function of $M$ in Fig.~\ref{fig:experimental_scaling} when learning the crosstalk matrix with our protocol.
The frequency error generally decreases with the training set size and reaches a plateau at $M=50$ with a median error of approximately $\SI{288}{kHz}$.

\section{Protocol scaling}

\begin{figure}[hbt!]
\subfloat{\label{fig:scaling}}
\subfloat{\label{fig:scaling_constant_M}}
\includegraphics{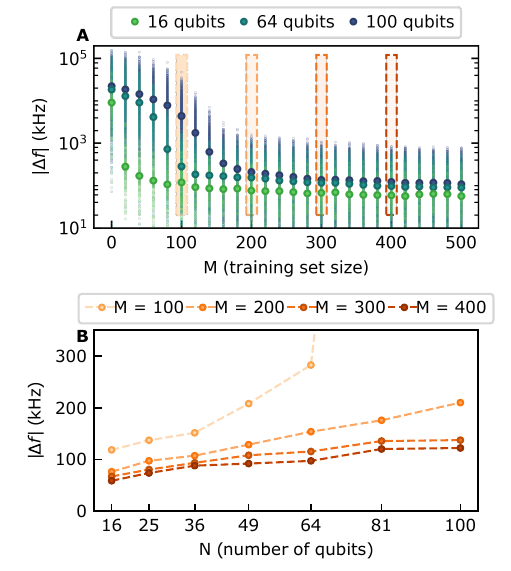}
\caption{
\textbf{Simulation of extensibility}
\textbf{(A)} Frequency error scaling for $N=16,64,100$ qubits.
We observe rapid convergence of frequency error after the training set size exceeds the number of qubits.
The distribution over $10$ different random realizations is shown using small circles, and the median values are shown using large circles.
\textbf{(B)} Vertical slices of the data, highlighted in \textbf{(A)} by shaded boxes, show the median frequency error as a function of the number of qubits for fixed training set size $M \in \{100,200,300,400\}$. 
The simulation assumes each qubit frequency measurement includes a measurement uncertainty of $\SI{0.5}{MHz}$.
}
\label{fig:Scale}
\end{figure}

In order to demonstrate the extensibility of our approach, we simulate the learning procedure with an extension of the crosstalk model (see Appendix~\ref{appendix:crosstalk_simulation_model}) for larger transmon array sizes.
We show the scaling of the frequency error for qubit arrays of size $N=16,\, 64,\,  \mathrm{and}\, 100$ qubits in Fig.~\ref{fig:scaling}, assuming conservatively a frequency measurement uncertainty of $\SI{0.5}{MHz}$.
We observe a rapid convergence of the error in target frequency when using a training set with a size larger than the number of qubits. 

In Fig.~\ref{fig:scaling_constant_M}, we study the error scaling by keeping the training set size constant (vertical slices of Fig.~\ref{fig:scaling}).
For training set sizes of $M=200, \,300, \,\textrm{and} \,400$, the frequency error empirically scales linearly with the number of qubits and remains below $\SI{200}{kHz}$. The sources of error used in our model simulations are inaccuracies in characterizing the transmon spectrum parameters and uncertainty in qubit frequency measurements. Hence, we expect the frequency error to be lower in simulation compared to our experiments.

Based on the scaling observed in the simulation, in addition to the experimental demonstration, we conclude that using our learning-based crosstalk optimization protocol we would be able to accurately train the crosstalk matrix with $M \approx 2 N$ spectroscopic frequency measurements per qubit.
Since we can perform simultaneous frequency measurements, the total measurement time for the training set will be proportional to $2 N$.

\section{Discussion}

In this work, we describe an extensible approach for accurate flux crosstalk characterization and calibration based on machine learning.
We experimentally verify the performance of our approach by employing the protocol to calibrate the static flux crosstalk of a 16-qubit flux-tunable transmon processor and observe convergence to a median frequency error less than $\SI{300}{kHz}$ with only $M=50$ spectroscopy measurements of each qubit frequency, which can be done in parallel for all qubits.
The protocol can also be used to calibrate the crosstalk matrix for fast flux pulses (see Appendix~\ref{appendix:Fast_flux_crosstalk_calibration}). 

We have demonstrated that our protocol enables us to realize a qubit frequency layout with accuracies better than $<\SI{300}{kHz}$. In order to perform high-fidelity operations, we measure the frequency of each qubit using Ramsey interferometry after setting the qubit frequencies. In Appendix~\ref{appendix:gate_error_analysis}, we discuss the impact of frequency errors on gate fidelities.
The learned crosstalk matrix from our protocol also extrapolates to frequency layouts outside of the crosstalk training region defined in Fig.~\ref{fig:transmon_spectrum} (see Appendix~\ref{appendix:other_bias_regions}).

The accuracy of the protocol for each qubit relies on a precise characterization of the transmon spectrum fit parameters.
Furthermore, spectrum irregularities caused by TLS defects coupled to the qubit inhibit the convergence of the crosstalk optimization.
To address the former, we perform simulations (see Appendix~\ref{appendix:Error_analysis}) that show the method is robust against realistic errors in each fit parameter.
To combat the impact of TLSs, we could use exclusion zones to avoid choosing frequencies in the spectrum that diverge from the transmon model.

We also study the performance of our protocol for different levels of crosstalk (Appendix~\ref{appendix:crosstalk_level_scaling}).
We find that our method can effectively learn the device crosstalk when off-diagonal elements stay below roughly $10\%$.
Flux crosstalk in current planar arrays of flux-tunable transmons is generally comfortably within this bound~\cite{abrams_methods_2019, yan_qwalk_2019, karamlou_quantum_2022, braumuller_probing_2022, zhang_bandgap_2023}, and it is even lower in multilayer devices that can better suppress crosstalk (see Fig.~\ref{fig:xtalk_matrix}).
This also suggests that routine recalibration of $\boldsymbol{S}$ will be efficient, and we numerically find it should require fewer measurements than the initial calibration (see Appendix~\ref{appendix:recalibration}).

Advances in calibration efficiency and extensibility are essential as processor sizes increase.
The learning-based flux crosstalk calibration procedure introduced in this work requires relatively few spectroscopic measurements, is robust to measurement error, and scales favorably as array size increases.
Alternatively, one can use Ramsey measurements to determine the qubit frequencies, which is faster than spectroscopy for single-qubit frequency measurement.
Future implementations could use simultaneous spectroscopy measurements to initially learn the matrix, and then use Ramsey measurements to fine-tune the calibration if necessary.
This approach reduces the characterization time and improves the accuracy of the flux crosstalk matrix characterization---and ultimately the performance of algorithms and simulations run on superconducting qubit processors. 

\bibliography{refs.bib}

\vspace{0.2in}
\section*{Acknowledgments}
The authors are grateful to Patrick M. Harrington and Francisca Vasconcelos for fruitful discussions. 
This work is supported in part by the U.S. Department of Energy, Office of Science, National Quantum Information Science Research Centers, Quantum System Accelerator (QSA); in part by the Defense Advanced Research Projects Agency under the Quantum Benchmarking contract; in part by U.S. Army Research Office Grant W911NF-18-1-0411; and in part by the Office of the Director of National Intelligence (ODNI), Intelligence Advanced Research Projects Activity (IARPA), the Department of Energy, and the Under Secretary of Defense for Research and Engineering under Air Force Contract No. FA8702- 15-D-0001. 
CNB acknowledges support from the STC Center for Integrated Quantum Materials, NSF Grant No. DMR-1231319 and the Wellesley College Samuel and Hilda Levitt Fellowship. 
AHK acknowledges support from the NSF Graduate Research Fellowship Program. 
SEM is supported by a NASA Space Technology Research Fellowship. 
ITR is supported by an appointment to the Intelligence Community Postdoctoral Research Fellowship Program at the Massachusetts Institute of Technology administered by Oak Ridge Institute for Science and Education (ORISE) through an interagency agreement between the U.S. Department of Energy and the Office of the Director of National Intelligence (ODNI). 
Any opinions, findings, conclusions, or recommendations expressed in this material are those of the author(s) and do not necessarily reflect the views of the DOE, DARPA, ODNI, IARPA, or USDR\&E.

\appendix

\section{3D-Integrated, Flip-Chip Device}
\label{appendix:device}

Our experimental sample is an array of 16 flux-tunable transmon qubits fabricated using a flip-chip process~\cite{rosenberg_2017}.
Unlike a planar architecture, where all chip elements are mounted on the same surface, the flip-chip has two separate tiers which are stacked on top of each other.
The qubit tier houses the qubits and the interposer tier houses all other chip elements.
The benefits of the flip-chip design include decreased distances between control lines and the qubits they target and increased shielding between neighboring qubits, which significantly reduces overall crosstalk levels for DC flux control and fast flux pulses.
For further details about the device see~\cite{karamlou_2023}.

{\renewcommand{\arraystretch}{2}%
\begin{table}[ht]
    \centering
    \begin{tabular}{|>{\hspace{6pt}}l<{\hspace{6pt}}||>{\hspace{6pt}}l<{\hspace{6pt}}|}
        \hline
        Qubit parameter & Measured Value \\
        \hline \hline
        $f^\text{max}$ & $4.887 \pm 0.110$ (GHz) \\ \hline
        $V^{\Phi_0}$ & $29.2 \pm 2.7$ (V)  \\ \hline
        $d$ & $0.35 \pm 0.04$  \\ \hline
        $E_C/h$ & $196.1\pm5.2$ (MHz)  \\ \hline
        $|\Phi_{\mathrm{offset}}|$ & $19.7 \pm 5.9$ (m$\Phi_0$)  \\ \hline 
    \end{tabular}
    \caption{\textbf{Qubit parameters} for the 3D-integrated flip-chip device. Mean values plus or minus one standard deviation are reported. Deviations in parameter values are due to unintentional fabrication imperfections.}
    \label{tab:qubit_parameters}
\end{table}
}

The nearest neighbor coupling on this device is fixed at $J/2\pi = (5.89 \pm 0.4)$~MHz, measured at qubit frequencies of $\SI{4.5}{GHz}$. Before performing the learning-based protocol, we characterize the qubit-qubit couplings should be characterized. With the detuning scheme we use for training ($>200$~MHz detuning between nearest neighbor qubits and $>50$~MHz detuning between any two qubits), we have a mean dispersive shift of around 180 kHz, which we correct for (see Appendix~\ref{appendix:freq_error_budget}). This shift is less than the final frequency setting precision using the optimized crosstalk matrix, which is $\approx \SI{300}{kHz}$. In our simulations, we incorporate frequency measurement error and observe frequency error convergence for $\sigma_\mathrm{meas.} = \SI{0.5}{MHz}$. We, therefore, expect that this relatively small dispersive shift does not greatly impact the final frequency setting precision of our protocol.

\section{Comparison to Direct Measurement Approach}
\label{appendix:direct_measurement}

One alternative to the learning-based approach to crosstalk calibration described in this paper is to directly measure each element of $\boldsymbol{S}$.
A downside to this approach is that the number of elements of the crosstalk matrix scales quadratically with the number of qubits in the array. 

\begin{figure*}[hb]
\subfloat{\label{fig:DMA_a}}
\subfloat{\label{fig:DMA_b}}
\subfloat{\label{fig:DMA_c}}
\subfloat{\label{fig:DMA_d}}
\includegraphics[width=1.0\linewidth]{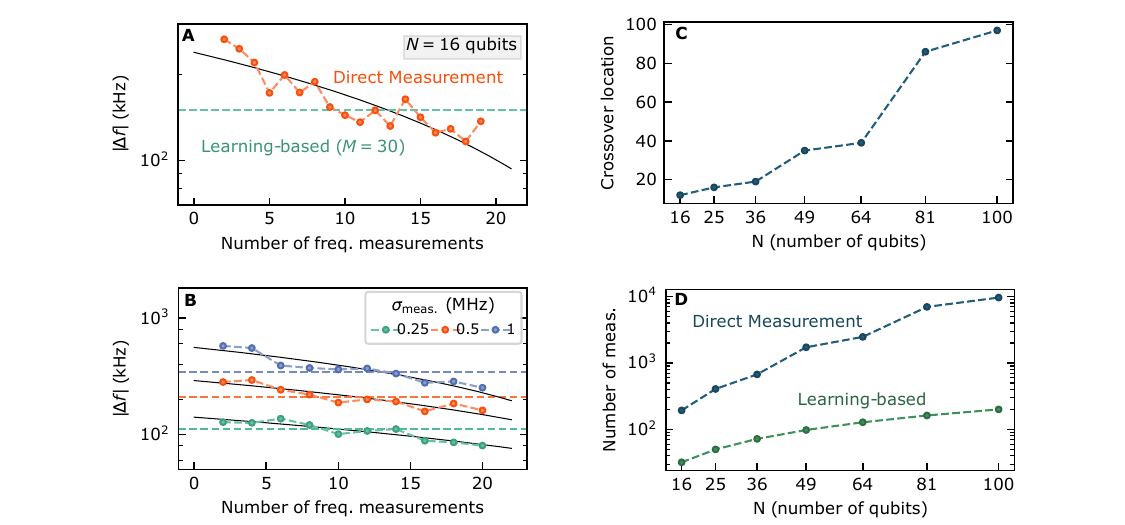}
\caption{
\textbf{Scaling of direct measurement approach}
Simulation of direct measurement convergence for a frequency measurement uncertainty of $\SI{0.5}{MHz}$.
We sweep the number of frequency measurements between $V_j = -0.3 V^{\Phi_0}$ and $V_j = +0.3 V^{\Phi_0}$.
\textbf{(A)} For $N=16$ qubits, the performance of the direct measurement approach exceeds the performance of the learning-based approach with $M=30\approx2\times N$ training sets when the number of frequency measurements is $\approx 10$.
The median frequency errors for the direct measurement approach are fitted linearly with a black line, which crosses over the learning-based median frequency error at $\approx 12$ data points.
The medians are computed over 10 simulation repetitions.
In \textbf{(B)}, we examine the performance of the direct measurement approach for $N=16$ qubits under varying levels of frequency measurement uncertainty $\sigma_{\mathrm{meas.}}$.
We find that the crossover location (again, for the median frequency error using the learning-based approach with $M=30\approx2\times N$ training sets) increases as $\sigma_{\mathrm{meas.}}$ increases.
The medians are computed over 20 simulation repetitions.
In \textbf{(C)}, we examine the scaling of this crossover location for up to $N=100$ qubits, assuming $\sigma_{\mathrm{meas.}}=\SI{0.5}{MHz}$.
The number of frequency measurements required to match the learning-based protocol's performance grows empirically linearly with the number of qubits.
\textbf{(D)} The number of frequency measurements required for each qubit, given by the crossover location times $N-1$, diverges from the learning-based $M=2\times N$ measurements per qubit.}
\label{fig:DMA}
\end{figure*}

We note that in our experimental setup, we can simultaneously measure all 16 qubits' frequencies, and we are able to leverage this simultaneous frequency measurement in the learning-based protocol. We cannot perform simultaneous frequency measurement in the direct measurement approach, because we cannot bias more than one qubit off of the sweet spot without introducing flux crosstalk effects. We might wish to sweep the voltage applied to one flux line and measure the response of all other qubits. Unfortunately, this approach to direct measurement will not work due to the broadness of the transmon spectrum (since $1/V^{\Phi_0}_{i,j} \ll 1/V^{\Phi_0}_{i,i}$ for $i\neq j$) and the relatively narrow tuning range for our voltages applied to flux lines. We will be unable to accurately fit the transmon spectrum and extract $V^{\Phi_0}_{i,j}$. To resolve elements of $\boldsymbol{S}$, we must first bias a qubit to a steep (flux-sensitive) section of the transmon spectrum and then sweep the voltage applied to another qubit's flux line.
Therefore, we cannot perform a simultaneous frequency measurement to obtain all the crosstalk elements due to one flux line.
There are creative methods by which we can obtain more than one crosstalk element via simultaneous measurements, but it will be impossible to obtain a full column of the crosstalk matrix at once in this fashion.

We simulate the direct measurement approach for a 16-qubit array while varying the frequency measurement uncertainty (as in Fig~\ref{fig:protocol_simulation}).
For the direct measurement approach, we bias qubit $i$ away from its sweet spot by applying a voltage $V_i$ to flux line $i$.
Then, we sweep the voltage of flux line $j$ across its full tuning range.
For our voltage source and qubit $V^{\Phi_0}$'s, this is approximately $\pm$0.3 times the average $V^{\Phi_0}$ of a qubit.
We fit the transmon spectrum, holding $f^{\text{max}}$, $E_{\text{C}}$, and $d$ fixed.
The new flux offset will be approximately:
\begin{equation}
    \Phi_{\text{offset}} = \frac{V_i}{V_{i,i}^{\Phi_0}} + \Phi_{\text{offset}, i}
\end{equation}
Using this initial guess for the flux offset and using the slope of the measured frequencies to determine the sign of $V_{i,j}^{\Phi_0}$ (to constrain it), we fit the curve to find $V_{i,j}^{\Phi_0}$ (we also obtain $\Phi_{\text{offset}}$ from the fit, but this information is irrelevant).
Finally, the measured element of the crosstalk matrix is:
\begin{equation}
    \boldsymbol{S}_{i,j} = \frac{V_{i,i}^{\Phi_0}}{V_{i,j}^{\Phi_0}}
\end{equation}

We see that for a reasonably assumed measurement uncertainty of $\SI{0.5}{MHz}$, the direct measurement approach reaches the same level of precision as the learning-based approach with a size $M=30$ training set when the number of data points reaches 10 (Fig.~\ref{fig:DMA_a}.
This tells us that the direct measurement approach requires 5x as many measurements to reach the same level of precision as the learning-based approach.

In the learning-based approach, each qubit's frequency is measured 30 times.
In the direct measurement approach, each qubit's frequency is measured 10 times for each flux line, for a total of $15\cdot 10 = 150$ times.
So for 16 qubits, we already achieve a $\approx 5\times$ speedup by using the learning-based protocol.

\section{Gradient Descent Optimizers}
\label{appendix:optimizers}

\begin{figure*}[ht]
\subfloat{\label{fig:GDO_a}}
\subfloat{\label{fig:GDO_b}}
\subfloat{\label{fig:GDO_c}}
\subfloat{\label{fig:GDO_d}}
\includegraphics[width=1.0\linewidth]{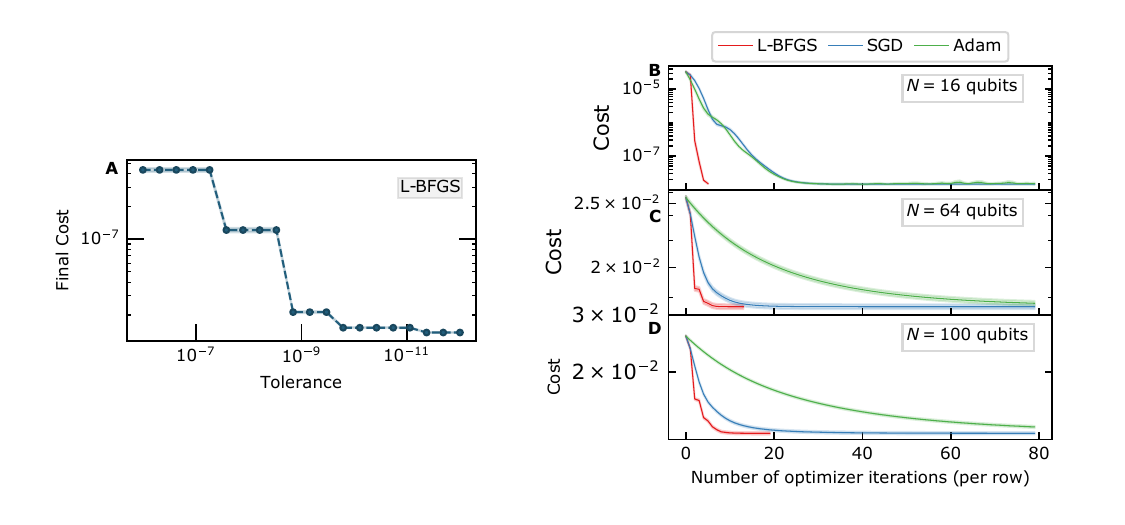}
\caption{
\textbf{Comparison of gradient descent optimizers}
We minimize a mean squared error cost function to optimize $\boldsymbol{S}$ row-by-row.
In \textbf{(A)}, we consider the final minimized cost versus the function change tolerance of the L-BFGS optimizer for $N=16$ qubits.
The mean costs were computed over 100 simulation repetitions, and the x-axis is inverted for clarity.
The shaded region is the 95\% confidence interval for the mean
In \textbf{(B,C,D)}, we compare the minimization of the cost function of the L-BFGS, SGD, and Adam optimizers for $N\in\{16,64,100\}$ qubits.
L-BFGS terminates on tolerance, whereas SGD and Adam terminate after a given number of iterations.
The L-BFGS tolerance is set to $10^{-10}$ with a learning rate of $1.0$.
The SGD learning rate is $1.0$ with a momentum of $0.7$.
The Adam learning rate is $0.002$ with betas of $(0.7, 0.999)$.
The SGD and Adam parameters were selected for the best performance for $N=16$ qubits.
The mean costs were computed over 10 simulation repetitions, and the shaded region is the 95\% confidence interval for the mean.
}
\label{fig:optimizers}
\end{figure*}

One critical piece of the protocol is minimizing the mean squared error cost function in a gradient descent optimizer.
Throughout this work, we use the L-BFGS optimizer in \textsf{PyTorch}, with a learning rate of $1.0$.
In order to ensure proper convergence of our gradient descent minimization, we compared the L-BFGS results to a couple of other \textsf{PyTorch} optimizers: SGD and Adam (see Fig.~\ref{fig:optimizers}).
On a practical level, it doesn't matter which optimizer we use, since this is not a bottleneck in our calibration protocol.
However, we find that L-BFGS converges with fewer optimizer iterations in comparison to other optimizers.
Additionally, other optimizers have additional parameters that need to be set.
For example, SGD has a momentum parameter that may need to be changed for optimal performance.

We find that L-BFGS consistently converges in the least iterations compared to other optimizers, and additionally requires no tuning of parameters, making it a good choice for practical use in the laboratory.

\section{Schematic of Learning-based Protocol}
\label{appendix:flowchart}

The goal of flux crosstalk calibration is to control qubit frequencies precisely.
The learning-based approach described in this paper is an intuitive approach to this problem since it actively addresses the goal of the calibration via the calibration process.
The intuitiveness of the learning-based approach stands in contrast to a direct measurement approach, which seeks to resolve individual elements of the crosstalk matrix $\boldsymbol{S}$, but does not directly evaluate the ability of the calibrated $\boldsymbol{S}$ to set qubit frequencies precisely.
For a quantitative comparison of these two approaches, see Appendix~\ref{appendix:direct_measurement}. 

\begin{figure*}[ht]
\includegraphics[width=1.0\linewidth]{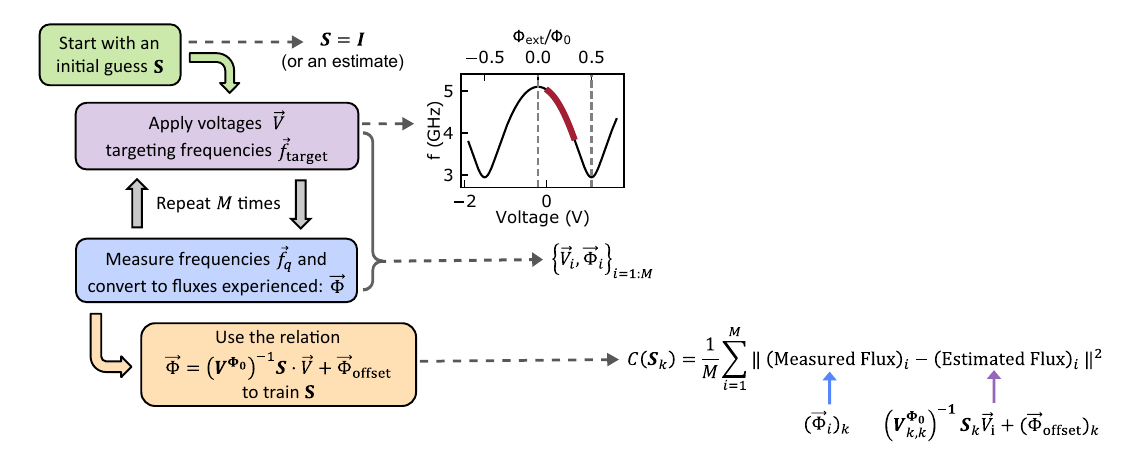}
\caption{
\textbf{Flowchart for learning-based protocol}
 We begin with an initial $\boldsymbol{S}$, which could be the identity or an estimate.
 We select target frequencies from a range of $\approx100~$MHz to $\approx1$~GHz off the sweet spot and apply voltages $\vec{V}$ targeting these frequencies.
 We measure the qubits' frequencies and convert them to fluxes experienced.
 After repeating this process $M$ times, we have a size-$M$ training set of pairs of applied voltages and experienced fluxes.
 We use the linear flux crosstalk relation to train each row of $\boldsymbol{S}$ $(\boldsymbol{S}_k)$ by minimizing the mean-squared-error cost function $C(\boldsymbol{S}_k)$ of the measured flux and the estimated flux in a gradient descent optimizer.
 The resultant $\boldsymbol{S}$ minimizes the difference between measured and experienced fluxes.}
\label{fig:flowchart}
\end{figure*}

The general concept behind the learning-based approach to flux crosstalk calibration is to use an initial estimate for $\boldsymbol{S}$ to target frequencies $\vec{f}_{\mathrm{target}}$ and then measure the frequencies of the qubits.
The difference between the measured and targeted frequencies gives us insight into how to adjust $\boldsymbol{S}$ to minimize frequency error.
Specifically, we utilize the linear flux crosstalk relation (Eq.~\ref{eq:crosstalk_eq}) and the transmon spectrum (Eq.~\ref{eq:transmon_spectrum}) to obtain a cost function with a well-defined gradient which can be minimized in a gradient descent optimizer.
In Fig.~\ref{fig:flowchart}, we provide a flowchart visualization of the calibration process.

We note that the cost function we are minimizing is convex due to the constrained frequency range for target frequencies.
Although the transmon spectrum is periodic, we restrict our target frequencies to one portion of the spectrum, allowing for a $100$~MHz buffer from the upper sweet spot (and an even larger buffer from the lower sweet spot).
With this restriction, assuming reasonable crosstalk levels of $<10\%$, we eliminate the possibility of converting the measured frequency to an incorrect experienced flux.

For the purposes of this paper, which are to introduce and demonstrate the efficacy and extensibility of the learning-based approach, we did not optimize end-to-end calibration time.
One way the efficiency of this calibration protocol could be improved is to learn the crosstalk matrix iteratively.
We could easily modify this protocol to learn $\boldsymbol{S}$ in batches.
Instead of applying all voltages and measuring all frequencies, we can instead perform a size $m$ training set with $m<M$, and then use the resultant trained $\boldsymbol{S}$ as our new estimate for $\boldsymbol{S}$ for the next training set.
The advantage to this batched approach is that each iterative $\boldsymbol{S}$ will become closer and closer to the optimal $\boldsymbol{S}$.
Therefore, the actual qubit frequencies will be closer to the target frequencies in the training sets, which means we need to scan over a smaller range of frequencies to locate our qubits.
Since frequency measurements are the most time-consuming part of this protocol, minimizing the time per measurement will speed up the protocol.

\section{Crosstalk model}
\label{appendix:crosstalk_simulation_model}

\begin{figure}[hb]
\subfloat{\label{fig:xtalk_matrix}}
\subfloat{\label{fig:xtalk_vs_dist}}
\includegraphics[width=1.0\linewidth]{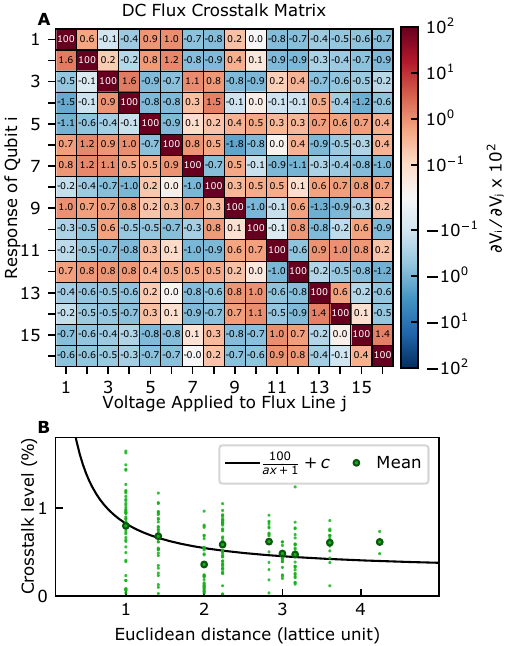}
\caption{
\textbf{The DC crosstalk matrix}
\textbf{(A)} The measured flux crosstalk sensitivity matrix for the 16-qubit flip-chip device in our lab, rescaled by a factor of 100 such that each element is a percentage.
The average off-diagonal crosstalk level is $<1\%$.
We use the distribution of crosstalk sensitivity $\boldsymbol{S}_{i,j}$ versus Euclidean distance between qubit $i$ and qubit $j$ to generate realistic crosstalk matrices for simulations of protocol scaling.
\textbf{(B)} The off-diagonal crosstalk versus Euclidean distance between flux line and qubit can be fitted with a $1/r$ decay defined by $\ell = \frac{100}{ax+1} + c $, with $a=178.2$ and $c=0.264$.}
\label{fig:xtalk_model}
\end{figure}

For all simulations of the calibration process for a 16-qubit array, we used the measured flux crosstalk matrix for the 16-qubit device discussed in this paper.
The crosstalk matrix for this device is shown in Fig.~\ref{fig:xtalk_matrix}.
The measured crosstalk matrix we used was from a previous cooldown, but the overall crosstalk matrix changes relatively little from one cooldown to the next.
For the simulation of protocol scaling, we needed a model for crosstalk to generate realistic crosstalk matrices for larger array sizes.

One possible way to model crosstalk is to consider the strength of crosstalk versus the Euclidean distance between the site of the flux line and the influenced qubit.
For the 16-qubit device, the magnitude of crosstalk vs Euclidean distance is shown in Fig.~\ref{fig:xtalk_vs_dist}. 

One compelling reason to model crosstalk in this way is that once current flows through a local flux line, it must flow through the superconducting ground plane to ground.
The paths that these currents take through the chip are unknown and cannot be modeled.
We do know, however, that a current dispensed at the end of a flux line must flow to the edge of the chip.
If we consider a circle of radius $r$ centered on the end of the flux line, we can assume there is an equal probability of the current flowing through any point on this circle (namely $s/2\pi r$ probability, for an arc of length $s$).
We can therefore assume that the probability of a current flowing past a point a distance $r$ from the end of a flux line is proportional to $1/r$.

We fit the curve in Fig~\ref{fig:xtalk_vs_dist} with a shifted $1/r$ decay: $\frac{1}{ax+1} + c $. This equation is a $1/r$ decay that passes through $(0,100\%)$ when $c=0$. We obtain fit values of $a=178.2$ and $c=0.264$.
The model qualitatively fits the data, with the nearest neighbor crosstalk slightly elevated compared to longer-range crosstalk. 

To generate crosstalk matrices for larger arrays, we used this exponential model.
Since we do not have experimental measurements for Euclidean distances beyond the furthest distance on a 16-qubit array, we used the standard deviation of the magnitude of all off-diagonal crosstalk elements for the 16-qubit array: $\sigma = 0.342$.
For a given crosstalk element $\boldsymbol{S}_{i,j}$ with $i\neq j$, we compute the crosstalk level $\ell$ corresponding to the Euclidean distance between qubit $i$ and qubit $j$ and then pick the magnitude of $\boldsymbol{S}_{i,j}$ from a normal distribution $\mathcal{N}(\ell, \sigma)$.
Then, we randomly multiply the element by $\pm 1$.

\section{Frequency Measurement Error Sources}
\label{appendix:freq_error_budget}

One of the limiting factors in the frequency precision of our protocol is the measurement error inherent in the training sets.
In each training set measurement, we measure the frequency of each qubit and convert that to a flux experienced.
The accuracy of this frequency measurement therefore directly influences the frequency error plateau for the protocol. 

{\renewcommand{\arraystretch}{2}%
\begin{table}[ht]
    \centering
    \begin{tabular}{|>{\hspace{6pt}}l<{\hspace{6pt}}||>{\hspace{6pt}}l<{\hspace{6pt}}|}
        \hline
        Error Source & Estimated Error \\
        \hline \hline
        Spectroscopic Inaccuracy & 120~kHz \\ \hline
        Uncompensated Dispersive Shift & 120~kHz  \\ \hline
        Residual ZZ Coupling & 230~kHz  \\ \hline\hline
        $\textbf{Total Estimate}$ & \textbf{280~kHz}  \\ \hline
    \end{tabular}
    \caption{\textbf{Error budget} for frequency measurement based on the known primary sources of uncertainty. Estimates for average error are reported.}
    \label{tab:error_budget}
\end{table}
}

We have identified three primary sources of error in our frequency measurements: spectroscopic measurement inaccuracies, uncompensated dispersive shifts, and residual ZZ couplings.
We report the estimated error from each of these sources in Table~\ref{tab:error_budget}.
In this section, we discuss these estimates and their implications. 

We measure our qubit frequencies via spectroscopy.
The spectroscopy scans have a frequency step of $\SI{0.5}{MHz}$ to rapidly scan $f_\mathrm{target} \pm \SI{30}{MHz}$ for each qubit.
We fit the peak to a Lorentzian lineshape. 
To obtain an estimate for the error of the fitted spectroscopy peak, we consider the distribution of fit uncertainties for the size $M=200$ initial training set used in our experiment.
After removing outliers where no peak was found, the mean error in the fitted peak is \SI{118}{kHz}. We report the estimated error from spectroscopic inaccuracies as \SI{120}{kHz}. 

The qubits experience dispersive shifts due to interactions with nearest neighbor qubits via direct capacitive coupling and with other qubits in the array via stray capacitances.
We characterize the qubit-qubit couplings before calibrating flux crosstalk, and we use these coupling strengths and the measured frequency of each qubit to determine the dispersive shift experienced by each qubit and compensate for it.
From the initial size $M=200$ training set in our experiment, the mean value of the calculated dispersive shift is approximately \SI{180}{kHz}.

Given a vector of measured frequencies $\vec{f}_\mathrm{meas.}$ and the qubit-qubit coupling matrix $\boldsymbol{J}$, we determine the uncoupled qubit frequencies $\vec{f}$ in the following way.
The elements of $\vec{f}$ are the frequencies of the qubits in the case of no qubit-qubit coupling: $\boldsymbol{J}=0$.
Therefore, we diagonalize the matrix $\boldsymbol{f} + \boldsymbol{J}$, where $\boldsymbol{f}$ is a diagonal matrix whose entries are $\vec{f}$.
The entries of this diagonalized matrix, which are the eigenvalues of $\boldsymbol{f} + \boldsymbol{J}$, will be $\vec{f}_\mathrm{meas.}$ (although perhaps in a different order).

To find $\vec{f}$, we first compute the eigenvalues $\vec{\lambda}$ of the matrix $\boldsymbol{f} + \boldsymbol{J}$ and sort them in ascending order.
We determine $\vec{f}$ by using a Nelder-Mead optimizer to minimize the cost function: 
\begin{equation}
    C(\vec{f}) = \bigg|\bigg| \vec{\lambda}(\vec{f}) - \vec{f}_\mathrm{meas.}\bigg|\bigg|^2
\end{equation}
where both $\vec{\lambda}(\vec{f})$ and $\vec{f}_\mathrm{meas.}$ are sorted in ascending order.

This dispersive shift compensation scheme relies on the accuracy of the characterized $\boldsymbol{J}$.
In practice, we do not characterize the full matrix $\boldsymbol{J}$, but instead, only measure nearest and next-nearest neighbor couplings.
Stray capacitances, however, can exist between any two qubits in an array.
Using the measured frequencies from the size $M=200$ initial training set, we compare the calculated $\vec{f}$ when using $\boldsymbol{J}$ versus using a modified $\boldsymbol{J}'$ with stray capacitances.
We set the beyond-next-nearest-neighbor elements of $\boldsymbol{J}'$ to the absolute value of random numbers normally distributed around zero, with a standard deviation of $\sigma \in J\times\{0.01,0.03,0.05\}$, where $J = 2\pi \times 5.98$~MHz is the mean nearest-neighbor coupling for our device.
The corresponding mean frequency error, assuming $\boldsymbol{J}'$ is the true coupling matrix, is $\Delta f =  49, 122, 164 $~kHz for $\sigma =0.01J,~0.03J,~0.05J$.
We therefore estimate that the error due to dispersive shift compensation is 120~kHz. This error is around 67\% of the average dispersive shift compensation of 180~kHz, which means that the overall benefits of our dispersive shift compensation are minimal, given our experimental conditions.

The qubits also have residual ZZ coupling, which leads to shifts in the qubit frequencies depending on the states of their neighbors.
Assuming the qubits have only three energy levels, the frequency shift due to residual ZZ coupling between two qubits (1 and 2) is given by:
\begin{equation}
    \Delta \omega_1 = \frac{\alpha_1 + \alpha_2}{(\Delta_{12} + \alpha_1)(\Delta_{12}-\alpha_2)}~J^2
\end{equation}
where $\alpha_i$ is the anharmonicity of qubit $i$, $J$ is the coupling strength between the qubits, and $\Delta_{12} = \omega_1 - \omega_2$ is the frequency detuning between the qubits \cite{malekakhlagh_2020,patterson_calibration_2019}.

From the measured frequencies of the size $M=200$ training set used in our experiment, we can find the mean shift due to nearest-neighbor ZZ coupling.
We assume that each anharmonicity is the mean of the anharmonicities: $\alpha_i/2\pi = -\SI{218.4}{MHz}$.
We also assume that each nearest neighbor coupling is the mean of the nearest neighbor couplings, $J/2\pi = \SI{5.89}{MHz}$, for our device, and the coupling between qubits that are not directly capacitively coupled is 0.
We find that the standard deviation of the frequency shifts due to ZZ coupling is approximately \SI{231}{kHz}.
Since next-nearest neighbor qubits also interact somewhat strongly and non-nearest neighbor qubits also interact via stray capacitances, we believe this estimate of error (\SI{230}{kHz}) to be conservative.

In Table~\ref{tab:error_budget}, we report the estimated error for each of these sources, as well as a total estimated error.
We assume these sources of error are uncorrelated and compute the total estimated error by adding in quadrature.
We estimate a total average error of \SI{280}{kHz}, which is less than the frequency measurement uncertainty of $\sigma_\mathrm{meas.} = \SI{0.5}{MHz}$ we use in our protocol simulations.
Other potential sources of frequency error are more challenging to characterize, such as small frequency shifts due to weak coupling with coherent defects or ZZ coupling due to higher transmon energy levels.
From the conservative error budget discussed above and other potential sources of error, we believe that $\sigma_\mathrm{meas.} = \SI{0.5}{MHz}$ is a reasonable upper bound for the frequency measurement uncertainty on our device.

\section{Fast flux crosstalk calibration}
\label{appendix:Fast_flux_crosstalk_calibration}

\begin{figure}[ht]
\includegraphics[width=1.0\linewidth]{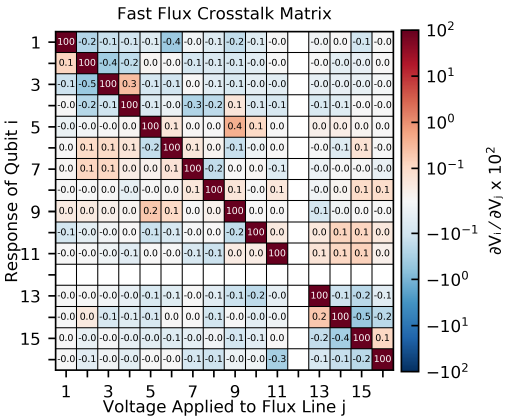}
\caption{
\textbf{The fast-flux crosstalk matrix}
The measured fast flux crosstalk matrix for the 16-qubit flip-chip device, rescaled by a factor of 100 such that each element is a percentage.
The average off-diagonal crosstalk level is approximately two orders of magnitude smaller for fast flux crosstalk than for DC flux crosstalk.
Elements labeled $\pm 0.0$ are crosstalk percentages of magnitude $< 0.05$\%.
The fast flux microwave line for qubit 12 was broken, so there is no information for this qubit.}
\label{fig:fast_flux}
\end{figure}

We can control our qubits via the local flux lines by applying DC voltages or baseband fast flux pulses. These fast flux pulses are square pulses with a cosine ramp and no modulations.
The learning-based protocol described in this paper can also be used for fast flux crosstalk calibration.
In Fig~\ref{fig:fast_flux}, we report the fast flux crosstalk matrix for our 16-qubit flip-chip device.
We find that the off-diagonal crosstalk levels are much smaller for fast flux pulses, with most off-diagonal elements having a magnitude on the order of $0.01\%$.
The average off-diagonal DC flux crosstalk level (Fig.~\ref{fig:xtalk_matrix}) is around $1\%$, so the fast flux crosstalk levels are around two orders of magnitude smaller than DC flux crosstalk levels.

In order to learn the fast flux crosstalk for each qubit, we tune that qubit to a target frequency using a $\SI{100}{ns}$ fast flux pulse and measure the qubit frequency via spectroscopy. We then use the same pulse amplitude for the target qubit, apply flux pulses with random amplitudes through the other flux lines, and measure the change in the target qubit's frequency. The voltages applied and the changes in the frequency of the target qubit from the training set that we use for learning the fast flux crosstalk matrix are shown in Fig.~\ref{fig:fast_flux}.

\section{Single-qubit Gate Fidelity}
\label{appendix:gate_error_analysis}

In this paper, we report an experimental median frequency error on our 16-qubit device of $\approx300$~kHz after learning the crosstalk matrix with a size $M=50$ training set.
We now address the effect of this error on single-qubit gate fidelity.

We would like to be able to arrange the qubits in frequency layouts with high fidelity so that we can perform quantum simulations and/or computations. If qubit frequencies are off from their target frequencies by over $\SI{10}{MHz}$ (as is the case on our device before learning the crosstalk), then qubits may interact more or less strongly than intended. An important thing to note is that after setting qubits in a frequency layout, we would perform Ramsey measurements to further refine the qubit frequencies (i.e. ensure we are driving qubits on resonance). Also, we could repeat the crosstalk learning protocol for a specific frequency layout to further reduce frequency errors.

\begin{figure}[ht]
\includegraphics[width=1.0\linewidth]{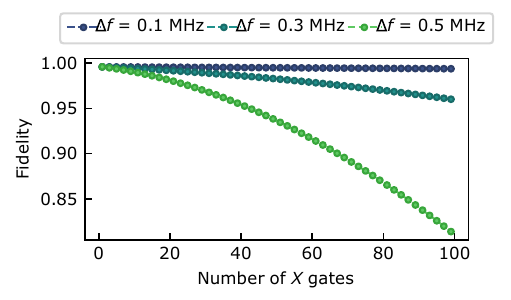}
\caption{
\textbf{Simulations of single-qubit gate fidelity}
in the presence of frequency error. The qubit is prepared in the excited state and then 2$\pi$-pulses are successively applied around the $X$-axis, in the presence of frequency detuning $\Delta f$. We assume a drive strength that induces a Rabi frequency of $\omega_R = 2\pi \cdot 10$~MHz.}
\label{fig:gate_error}
\end{figure}

In the event that we do not tune up our drive frequencies, we could consider the impact of the frequency error on our single-qubit gate fidelities. For an $X$-gate, a drive detuning tilts the effective driving field away from the $X$-axis in the $X-Z$ plane. In Fig.~\ref{fig:gate_error}, we can see the unitary fidelity for $\Delta f \in \{0.1, 0.3, 0.5\}$~MHz as we apply successive $2\pi$ rotations around the $X$-axis, with a pulse amplitude corresponding to a Rabi frequency of $\omega_R = 2\pi \cdot 10$~MHz. For $\Delta f < 0.3$~MHz, we maintain over 95\% fidelity with 100 successive $2\pi$-$X$ gates.
We therefore consider 300~kHz to be an acceptable median frequency error, especially since we can further improve upon this frequency error by performing Ramsey measurements or learning the crosstalk for a specific frequency layout.

\section{Bias Region Extrapolation}
\label{appendix:other_bias_regions}

We use the frequency region spanned by 100~MHz to 1~GHz off the sweet spot to learn our crosstalk matrix.
We now explore whether the crosstalk matrix learned in this region extrapolates to other bias regions.

\begin{figure}[ht]
\subfloat{\label{fig:bias_regions_a}}
\subfloat{\label{fig:bias_regions_b}}
\includegraphics[width=1.0\linewidth]{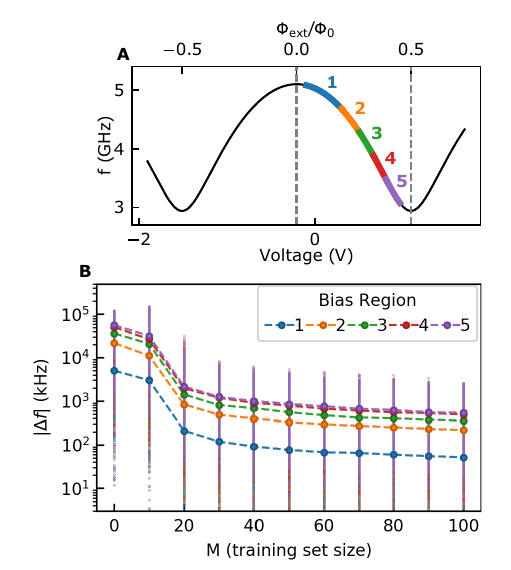}
\caption{
\textbf{Simulation of frequency error
in other bias regions.}
 The crosstalk is learned by placing the qubit frequencies randomly in the standard training region spanned by 100~MHz to 1~GHz off the sweet spot, and then the error in setting qubit frequencies is determined for 5 different bias regions, highlighted in \textbf{(A)}. We target random frequency vectors in each bias region to evaluate the learned crosstalk matrix. For these simulations, we disregard the detuning conditions normally used in our experiments.
 In \textbf{(B)}, we show the convergence of frequency error in each bias region as the training set size increases. The frequency precision is best for Region 1, closest to the sweet spot, and decreases the further the bias region is from the upper sweet spot.
 The medians are computed over 100 simulation repetitions, and all data points are shown on the plot.
}
\label{fig:bias_regions}
\end{figure}

In Fig.~\ref{fig:bias_regions}, we simulate the protocol's performance when evaluated in different bias regions. We still learn the crosstalk matrix using the standard training region (100~MHz to 1~GHz off the sweet spot), and then report the frequency error when attempting to set qubit frequencies to the 5 different bias regions highlighted in Fig.~\ref{fig:bias_regions_a}. For each bias region, we choose frequency vectors that place each qubit in that bias region (i.e. all voltages are nonzero). We note that in these simulations, we disregard the detuning conditions normally required in our experiments.

In Fig.~\ref{fig:bias_regions_b}, we see that the performance of the protocol is best for the bias region closest to the sweet spot, which makes sense since the applied voltages are the smallest. The performance of the protocol decreases the further the bias region is from the upper sweet spot, with the worst performance occurring in regions 4 and 5, with $\approx 800$~kHz and $\approx 850$~kHz median frequency error, respectively, for training set size $M=50$. 

In Appendix~\ref{appendix:gate_error_analysis}, we found that a frequency error of $500$~kHz impacts gate fidelities, particularly for more than 20 successive gates. Therefore, we expect that using the learned crosstalk matrix from the standard training region for frequency layouts with qubits primarily placed in bias regions 4 and 5 will result in worsened computational performance.

We note, however, that the frequency error in these regions follows the same convergence pattern as in other regions while starting from a higher initial error. In this sense, the learned crosstalk matrix extrapolates to different bias regions. Furthermore, we expect that frequency layouts with different qubits placed in different bias regions (as described in Fig.~\ref{fig:bias_regions_a}) will have a median frequency error somewhere in between the minimum and the maximum errors in Fig.~\ref{fig:bias_regions_b}.
To improve frequency setting precision in other bias regions, we would recommend repeating crosstalk training for the specific frequency layouts or bias regions desired. Additionally, we can perform Ramsey measurements to further refine qubit frequencies.

\section{Single-Qubit Characterization}
\label{appendix:single-qubit}

Before we can conduct the learning-based protocol described in this paper, we must perform single-qubit characterizations.
One of the essential assumptions of the protocol is that we can convert accurately between flux experienced by a qubit's SQUID loop and its frequency.
This conversion is described by the transmon spectrum (Eq.~\ref{eq:transmon_spectrum}).
In the case of a single qubit, the protocol relies on the ability to convert between voltage $V$ applied to its flux line and flux $\Phi_\mathrm{ext}$ experienced by the qubit's SQUID loop.
This conversion is described by $\Phi_\mathrm{ext} = V/V^{\Phi_0}+ \Phi_{\text{offset}}$, where $V^{\Phi_0}$ is the voltage required to tune the qubit by one magnetic flux quantum $\Phi_0$, and $\Phi_{\text{offset}}$ is a flux offset. 

We determine $V^{\Phi_0}$, $\Phi_{\text{offset}}$, $f^{\text{max}}$, $d$, and $E_{C}$ by performing qubit spectroscopy.
We sweep the voltage $V$ across the full extent of our tuning range (which in practice turns out to be approximately $\pm 0.3 V^{\Phi_0}$) and measure the qubit frequency.
We fit this curve with Eq.~\ref{eq:transmon_spectrum}, substituting $\Phi_\mathrm{ext} = V/V^{\Phi_0}+ \Phi_{\text{offset}}$, where $V^{\Phi_0}$.
Naturally, the more data points we take for the spectroscopy fit, the more precise our fit parameters will become.
We investigate the precision of our fit parameters under the presence of frequency measurement uncertainty $\sigma_{\mathrm{meas.}}\in \{0.1,0.5,1.0\}$~MHz. 

\begin{figure*}[ht]
\includegraphics[width=1.0\linewidth]{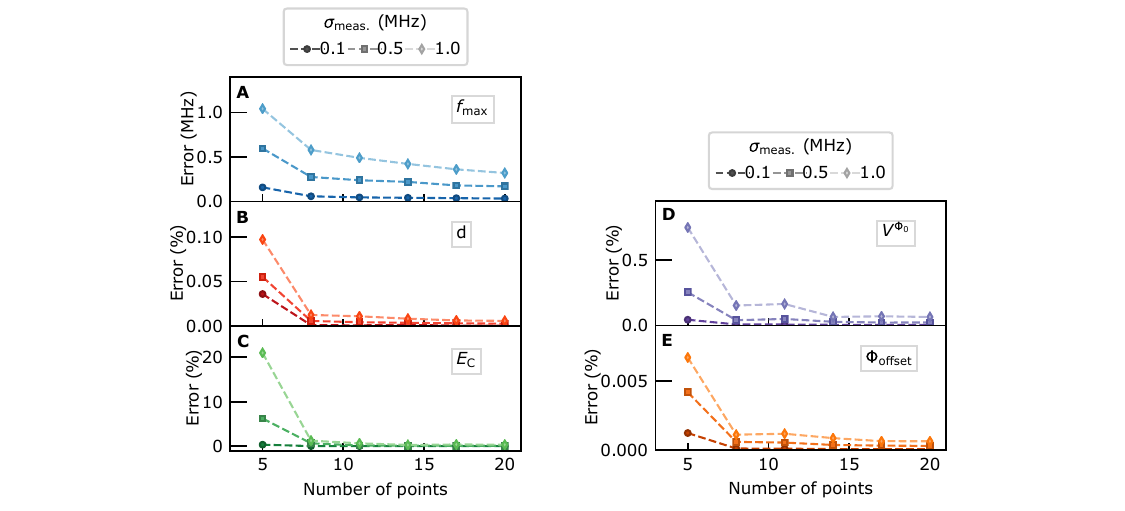}
\caption{
\textbf{Sensitivity analysis of transmon spectrum fits} The mean error in \textbf{(A)} $f^{\text{max}}$ (MHz), and the mean percent error in \textbf{(B)} $d$, \textbf{(C)} $E_{C}$, \textbf{(D)} $V^{\Phi_0}$, and \textbf{(E)} $\Phi_{\text{offset}}$, versus the number of data points in the spectroscopy sweep for varying levels of frequency measurement uncertainty $\sigma_{\mathrm{meas.}}$.
The error converges by around 15 measurements.
The means are computed over 50 simulation repetitions.}
\label{fig:Transmon_params_error}
\end{figure*}

In Section~\ref{appendix:Error_analysis}, we investigate how errors in fit parameters impact the convergence of the learning-based protocol.
From our analysis in this section (Fig.~\ref{fig:Transmon_params_error}), we can see that we can achieve the required precision in fit parameters with $<15$ frequency measurements per qubit.

\section{Error Analysis}
\label{appendix:Error_analysis}

We investigated, via simulation, the effect of error in the system on the efficacy of the protocol.
One key assumption of the protocol is that we can convert between a qubit's frequency $f$ and flux experienced by its SQUID loops $\Phi$ using the transmon spectrum, Eq.~\ref{eq:transmon_spectrum}.
If the parameters $f^{\text{max}}$, $d$, or $E_{C}$ have significant errors, the protocol will break down because we will be unable to accurately determine the fluxes experienced by the SQUIDs to train $\boldsymbol{S}$ (i.e. we won't have an accurate training set $\{\vec{V}_i, \vec{\Phi}^{\text{meas}}_i \}_{i=1:M}$).
We note that Eq~\ref{eq:transmon_spectrum} is an estimation, which is another potential source of error in our experimental results.
We assume, however, that the transmon spectrum is an exact equation for all simulations of protocol performance.

\begin{figure*}
\includegraphics[width=1.0\linewidth]{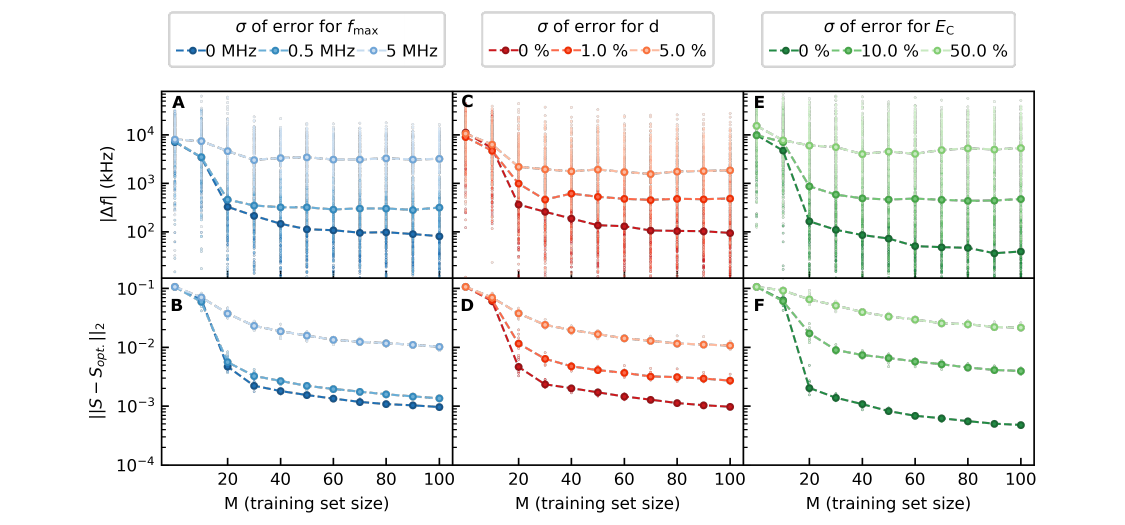}
\caption{
\textbf{Effect of error in transmon spectrum parameters on crosstalk precision}
Simulation of protocol convergence for varying levels of error in $f^\text{max}$ \textbf{(A,B)}, $d$ \textbf{(C,D)}, and $E_c$ \textbf{(E,F)}.
The frequency error is shown in \textbf{(A,C,E)} and the Euclidean norm between the trained crosstalk matrix $S_\text{opt.}$ and the actual crosstalk matrix $\boldsymbol{S}$ is shown in \textbf{(B,D,F)}.
The distribution over $10$ different random realizations is shown using small circles, and the median values are shown using large circles.
The convergence of the protocol is robust to small errors in these parameters but breaks down in the large error limit.
The medians are computed over 10 simulation repetitions, and all data points are shown on the plot.
A frequency measurement uncertainty of $\SI{0.5}{MHz}$ (see Fig.~\ref{fig:protocol_simulation}) was used for these simulations.}
\label{fig:ParamErr}
\end{figure*}

We simulated the calibration protocol for the 16-qubit array while varying the standard deviation of error for these parameters, shown in Fig.~\ref{fig:ParamErr}.
We find that the protocol begins to fail when the standard deviation of $f_\text{max}$ exceeds $\SI{0.5}{MHz}$, the standard deviation of $d$ exceeds 1\%, or the standard deviation of $E_C$ exceeds 10\%.

Another key assumption of the protocol is that we can form an accurate estimate for the flux experienced by the SQUID loops, given an optimized crosstalk matrix and the voltages applied.
This is the linear crosstalk relation:
\begin{equation}
    \vec{\Phi}_\mathrm{ext} = {(\boldsymbol{V}^{\Phi_0})^{-1}\boldsymbol{S}}\vec{V} + \vec{\Phi}_{\text{offset}}
    \label{eq:crosstalk_eq_repeated}
\end{equation}
This assumption breaks down when there are errors in the measured $\boldsymbol{V}^{\Phi_0}$ and $\vec{\Phi}_\text{offset}$.
Errors in $\boldsymbol{V}^{\Phi_0}$ and $\vec{\Phi}_\text{offset}$ will render the argument of the cost function inaccurate:
\begin{equation}
    C(\boldsymbol{S}_k) = \frac{1}{M} \sum\limits_{i=1}^M \left|\left| (\vec{\Phi}_{i})_k - \left[ {(\boldsymbol{V}^{\Phi_0}_{k,k})^{-1}\boldsymbol{S}_k} \vec{V}_i + (\vec{\Phi}_{\text{offset}})_k \right]\right|\right|^2
\label{eq:cost_function}
\end{equation}
preventing us from converging to the correct minimum.

\begin{figure*}[ht]
\subfloat{\label{fig:subplot_a}}
\includegraphics[width=1.0\linewidth]{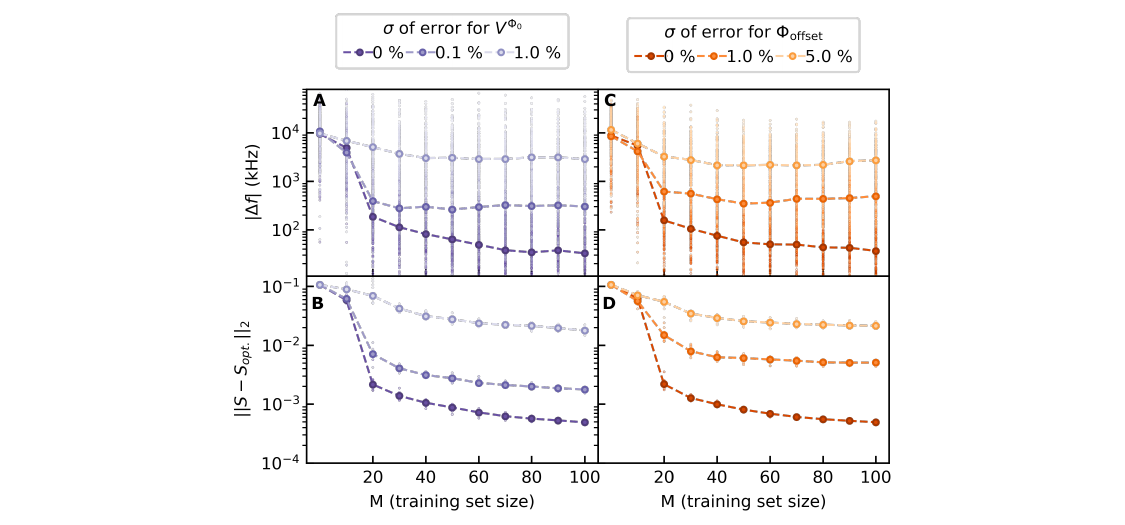}
\caption{
\textbf{$\boldsymbol{V}^{\Phi_0}$ and $\Phi_\text{offset}$ error analysis}
Simulation of protocol convergence for varying levels of error in $\boldsymbol{V}^{\Phi_0}$  \textbf{(A,B)} and $\Phi_\text{offset}$ \textbf{(C,D)}.
The frequency error is shown in \textbf{(A,C)} and the Euclidean norm between the trained crosstalk matrix $S_\text{opt.}$ and the actual crosstalk matrix $\boldsymbol{S}$ is shown in \textbf{(B,D)}.
The distribution over $10$ different random realizations is shown using small circles, and the median values are shown using large circles.
The convergence of the protocol is particularly sensitive to errors in $\boldsymbol{V}^{\Phi_0}$, and fairly sensitive to errors in $\Phi_\text{offset}$.
A frequency measurement uncertainty of $\SI{0.5}{MHz}$ (see Fig.~\ref{fig:protocol_simulation}) was used for these simulations.}
\label{fig:VOErr}
\end{figure*}

We simulated the calibration protocol for the 16-qubit array while varying the standard deviation of error for these parameters, shown in Fig.~\ref{fig:VOErr}.
We find that the protocol fails when the standard deviation of $\boldsymbol{V}^{\Phi_0}$ exceeds $0.1\%$ or the standard deviation of $\vec{\Phi}_\text{offset}$ exceeds 1\%.

\section{Efficient Recalibration}
\label{appendix:recalibration}

An advantage to the learning-based approach to calibration is that it can be used to efficiently recalibrate the crosstalk matrix.
With a direct measurement approach, there is no simple way to recalibrate the crosstalk without repeating the measurements over again.

\begin{figure}[ht]
\subfloat{\label{fig:RC_a}}
\subfloat{\label{fig:RC_b}}
\includegraphics[width=1.0\linewidth]{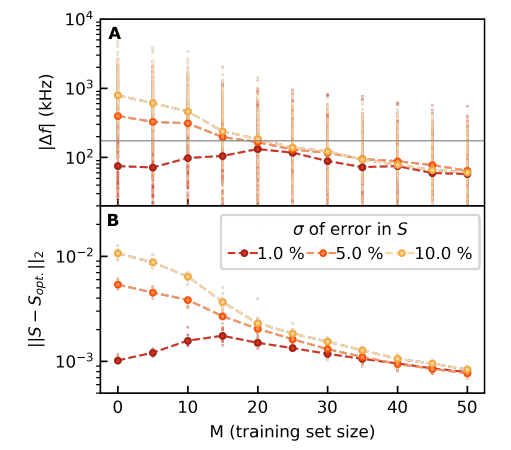}
\caption{
\textbf{Crosstalk matrix recalibration}
Simulation of protocol convergence for varying levels of error in the off-diagonal estimates of the initial crosstalk matrix used at the start of training.
Both the \textbf{(A)} median frequency error and the \textbf{(B)} median Euclidean norm between the trained crosstalk matrix $\boldsymbol{S}_\text{opt.}$ and the actual crosstalk matrix $\boldsymbol{S}$ converge by $M=20$ to the originally trained median from Fig~\ref{fig:protocol_simulation} (grey line, assuming $M=30$ and measurement uncertainty = $\SI{0.5}{MHz}$).
The medians are computed over 10 simulation repetitions, and  all data points are shown on the plot.
A frequency measurement uncertainty of 0.25~MHz was used for these simulations.
Since we already know where the qubits are, we can assume a finer frequency scan (lower power leading to less power broadening) and therefore less measurement uncertainty.}
\label{fig:recalibration}
\end{figure}

We follow the same procedure, except we use a good initial guess for $\boldsymbol{S}$ instead of the identity.
This will mean that the qubit frequencies will already be close to their targets with no additional training, enabling us to perform a finer spectroscopy scan with less measurement uncertainty.
In Fig~\ref{fig:recalibration}, we can see that the frequency error quickly converges to the median error from Fig~\ref{fig:protocol_simulation} (grey line, assuming $M=30$ and measurement uncertainty = $\SI{0.5}{MHz}$).
This means that we can use the same protocol to recalibrate our crosstalk matrix efficiently.

\section{Training \texorpdfstring{$V^{\Phi_0}$}{Lg}  and \texorpdfstring{$\Phi_{\mathrm{offset}}$}{Lg}}
\label{appendix:retrain_vo_offset}

Another advantage of this learning-based protocol is that it can be used to efficiently retrain $\boldsymbol{V}^{\Phi_0}$ and $\vec{\Phi}_{\mathrm{offset}}$.
For our superconducting devices, the DC flux crosstalk matrix remains essentially constant throughout an entire cooldown and also changes minimally from cooldown to cooldown.
The qubits' $V^{\Phi_0}$ and $\Phi_{\mathrm{offset}}$, however, can drift throughout a cooldown.
In particular, the flux offsets can drift substantially.

Once we have learned $\boldsymbol{S}$, we can use the same learning-based protocol to retrain $\boldsymbol{V}^{\Phi_0}$ and $\vec{\Phi}_{\mathrm{offset}}$.
Instead of optimizing $\boldsymbol{S}$ in the gradient descent optimizer, we can instead optimize one or both of $\boldsymbol{V}^{\Phi_0}$ and $\vec{\Phi}_{\mathrm{offset}}$ by minimizing $C(\boldsymbol{V}^{\Phi_0}, \vec{\Phi}_{\mathrm{offset}})$.

\section{Crosstalk level scaling}
\label{appendix:crosstalk_level_scaling}

The calibration approach outlined in this paper is applicable to other systems with low to moderate levels of crosstalk.
We simulated the convergence of the protocol for a 16-qubit array while varying the magnitude of the off-diagonal crosstalk elements of the target crosstalk matrix $\boldsymbol{S}_\text{target}$.
For a given off-diagonal crosstalk level $\ell$, we randomly assign each off-diagonal element of $\boldsymbol{S}_\text{target}$ to be $\pm \ell$.
In each simulation repetition, this arrangement of plus and minus changes, but the magnitude of each off-diagonal element remains fixed.

\begin{figure*}[ht]
\subfloat{\label{fig:XTL_a}}
\subfloat{\label{fig:XTL_b}}
\subfloat{\label{fig:XTL_c}}
\subfloat{\label{fig:XTL_d}}
\includegraphics[width=1.0\linewidth]{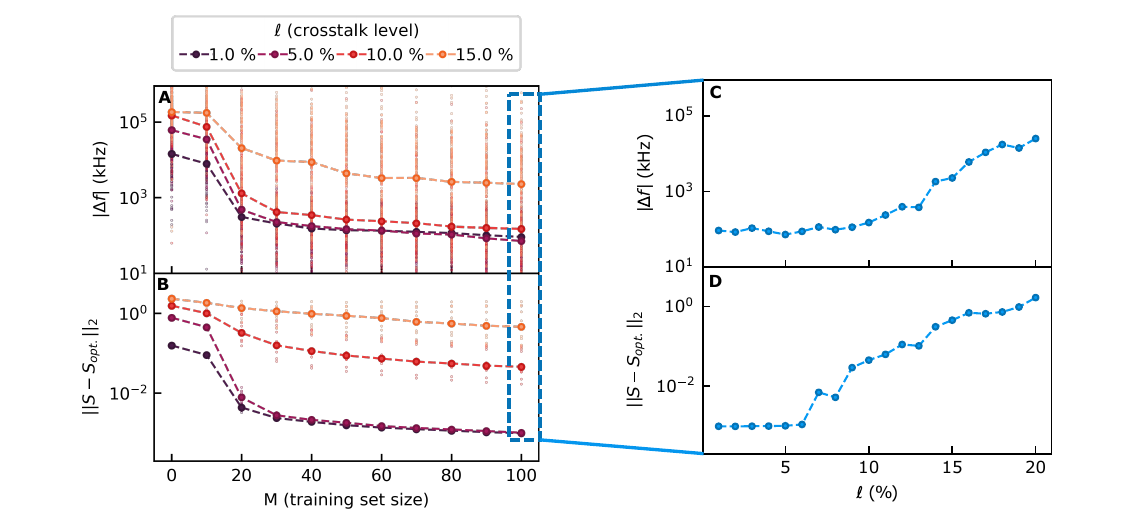}
\caption{
\textbf{Off-diagonal crosstalk level}
Simulation of protocol convergence for varying levels of off-diagonal crosstalk in the target crosstalk matrix $\boldsymbol{S}_\text{target}$.
Both the median frequency error \textbf{(A)} and the median Euclidean norm between the trained crosstalk matrix $S_\text{opt.}$ and the actual crosstalk matrix $\boldsymbol{S}$ \textbf{(B)} begin to fail to converge when off-diagonal crosstalk levels exceed 10\%.
For a training set of size $M=100$, the median frequency error \textbf{(C)} and median crosstalk matrix error \textbf{(D)} grow exponentially for large off-diagonal crosstalk levels.
The medians are computed over 10 simulation repetitions, and all data points are shown on the plot.
A frequency measurement uncertainty of $\SI{0.5}{MHz}$ (see Fig.~\ref{fig:protocol_simulation}) was used for these simulations.}
\label{fig:XTL}
\end{figure*}

We find that the protocol breaks down when the off-diagonal crosstalk levels exceed 10\%.
For a training set of size $M=100$, the median crosstalk matrix error Fig.~\ref{fig:XTL_d} begins to rise after off-diagonal crosstalk levels reach 6\%.
While crosstalk matrix error is a potentially useful metric for protocol simulations, it is not measurable experimentally, as the target crosstalk matrix is unknown.
The experimentally measurable quantity, median frequency error Fig.~\ref{fig:XTL_c}, begins to rise later after off-diagonal crosstalk levels reach 10\%.
This protocol will function well for systems with crosstalk levels $\leq 10\%$.

\end{document}